# The Species Abundances Distribution in a new perspective

M. Ravasz[1], A. Balog[2], V. Markó[2] and Z. Néda[1,3]

E-mail: zneda@phys.ubbcluj.ro

[1] *Babeş-Bolyai University, Dept. of Physics, RO-40048 Cluj, Romania*
[2] *Corvinus University, Dept. of Entomology, HU-1118 Budapest, Hungary*
[3] *University of Notre Dame, Dept. of Physics, IN 46556-5670 Notre Dame, USA*

**Studies on distribution, abundance and diversity of species revealed fascinating universalities in macroecology [1,2]. Many of these patterns, like the species-area and range-abundance relationship [3,4,5] or the year-to-year fluctuations in population sizes [6] are expressed as power-law distributions, and indicate thus scale-invariance. The species abundance distribution (SAD) apparently shows this scale-free nature only for relatively rare species, and its mathematical form is much debated. In the present work we propose a new mathematical expression for SAD which describes reasonable well most of the presently available large-scale experimental data and the results of the neutral models [7]. This distribution function leads to an interesting relation between the total number of individuals, total number of species and the size of the most abundant species of the meta-community. This novel scaling relation is confirmed by computer simulations on neutral models.**

The species abundance distribution (SAD) is introduced for characterizing the frequency of species with a given abundance [8,9]. It is often studied for communities of ecologically similar species that compete with each other only for resources [7,10,11]. There are usually two ways to characterize species abundance [12,13]. One possibility is to rank all species after their abundance, and plot for each species their percentage in the whole community as a function of their rank. The second way of characterizing species abundance is by constructing a species abundance distribution histogram. Since there are usually small numbers of species with large sizes, the tail of a simple histogram would have large fluctuations. In order to get a smooth tail for this distribution, a logarithmic binning is considered [8], which means that we count the number, $S_k$, of species with sizes between $2^k$ and $2^{k+1}$ (*k=0,1,2,3 ….*), i.e. we construct the histogram on intervals that are not of constant length, but exponentially increasing. In the literature, SAD is than plotted as the logarithm of $S_k$ versus *k*. This curve is bell-shaped and it is believed to be a log-normal distribution [8], although it's actual form is still much debated [12]. It is often wrongly considered [10,14] that this bell-shaped histogram indicates that few species are either extremely common or rare, and most of the species are of moderate or relatively low abundance. Without questioning the relevance of this histogram in characterizing SAD, in this letter the rigorously defined distribution function is used and a simple one-parameter fit is given to approximate its mathematical form. The results of several experimental data are successfully fitted and it is shown that the proposed equation describes also well the SAD obtained from the much debated neutral model. Moreover, this new equation for SAD describes also the scaling-laws observed in computer simulations on neutral models.

Many distributions were used (for a review see [2] or [15]) for fitting SAD in different meta-communities. Nowadays it is believed that there is no magic formula that would describe in a general manner the species abundances for all meta-communities. The most used form to describe the shape of SAD is the so called log-normal distribution, which is supported by arguments based on the central limit theorem of independent random variables. Other distributions used to describe the shape of SAD are the uniform distribution, the broken-stick distribution, the negative binomial distribution, the geometric series distribution, the logseries distribution and the non-analytic zero-sum multinomial (ZSM) distribution. ZSM has no analytical form and it is generated numerically by neutral models [12].

As discussed by May [15], the rigorously defined distribution function for SAD should characterize the number of species with sizes between $x$, and $x+dx$, for a **unit $dx$** interval. This would mean that in the histograms that are usually considered for SAD, one must divide $S_k$ with the exponentially increasing size of the interval, which is $2^k$. If we want to be even more rigorous and work with a normalized distribution function $f(x)$, we also have to divide $S_k$ by the total number of species ($S$), but this would not change anymore the overall shape of the distribution function. The correct mathematical formula for $f(x)$ would be than $f(x)=S_x/(S.2^x)$. Constructing the distribution functions in this manner one will obtain a monotonically decreasing function, resembling a power-law with a negative exponent. For low abundances usually a power-law fit with an exponent -1 works reasonable well. A plot on a log-log scale reveals however that this simple power-law approximation for the whole abundances interval is inadequate, and significant deviations are observed for abundant species. More particularly, one finds that for high abundances the shape of the distribution function decreases more rapidly than a simple power-law. This effect is of course expected, since there is an obvious cut-off in the system, governed by the size of the considered habitat. Evidently, in a finite meta-community the number of individuals in the most abundant species $N_s$ is limited, and this introduces a cutoff in the distribution function. Analyzing several large datasets and also the results of the neutral models, we empirically found that a simple fit of the form

$$f(x) = C \frac{C_1 - x}{x} \tag{1}$$

describes reasonable well the whole shape of SAD. In equation (1) $C$ is a normalization constant which can be determined as function of $C_1$. The distribution function (1) suggests a scaling behavior for rare species, and a cutoff at $x=C_1$, which means that $C_1 \cong N_s$. Using the obvious

$$C \int_1^{N_s} \frac{N_s - x}{x} dx = 1, \tag{2}$$

normalization condition, we get the form of the normalized SAD:

$$f(x) = \frac{1}{N_s \ln(N_s) - N_s + 1} \frac{N_s - x}{x} \tag{3}$$

To argue in favor of the form (1) of SAD, first we fit several large experimental datasets. We re-plotted the publicly available results from the Barro Colorado Island (BCI) tree dataset [16-18], from the North American Breeding Bird Survey (BBS) database [19-21] and the light-trap measurements of Dirks [22], Seamans (unpublished data, results taken from [8]) and Williams [9] on moths (***Insecta: Lepidoptera***).

On the first graph on Figure 1 the BCI measurement results for SAD are re-plotted and fitted with equation (3). The measurements are on three different years and look very similar. All of them can be fitted acceptably by choosing the value of $N_s \cong 13000$. Similar results were obtained using the BBS data on the whole state of Alabama. Studying the survey for 1980 and 1990 one obtains again similar data, and the shape of SAD can be successfully approximated using $N_s \cong 4100$ (second graph on Figure 1). The light-trap measurements on moths are plotted and fitted on the last graph of Figure 1. For the independent measurements made by Dirks and Seamans the best fit parameters are $N_s \cong 1000$, and $N_s \cong 4500$, respectively. Very similar results can be obtained re-plotting the results of Willimas [9], however this graph is not shown since it would overcrowd the plot.

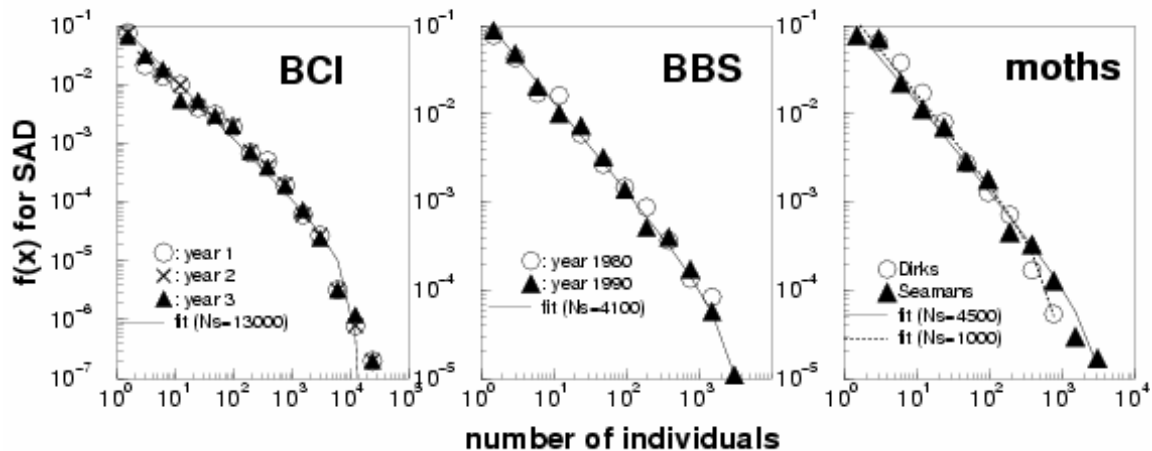

*Figure 1.* *Experimental results on SAD fitted by equation (3). Results for the BCI dataset on three different years (best fit parameter $N_s \cong 13000$ for all years), for the BBS dataset on the whole state of Alabama for 1980 and 1990 (best fit parameter $N_s \cong 4100$ for both years) and for moths as measured by Dirks and Seamans (best fit parameter $N_s \cong 1000$ and $N_s \cong 4500$, respectively)*

Computer simulations on neutral community models implemented on a lattice supports also the conjecture (1) for the form of SAD. For neutral models SAD is generated numerically and it is called the ZSM distribution [7,12,23]. We will argue thus that the ZSM distribution can be successfully fitted using the form (1). We use two different simulation codes, one elaborated by McGill to test the unified neutral theory of biodiversity [12] and one written by us.

In the model considered by us, **S** number of species can coexist on a **20x20** lattice, and all individuals from one species have the same probability per unit time for multiplication (**b**), death (**d**), and diffusion (**q**) to a nearby site. In the neutral version of the model all species are considered equally fit for the given ecosystem and have thus the same multiplication, death and diffusion rate. On each lattice site the total number of individuals is limited to $N_{max}$. Once the number of individuals on a site exceeds $N_{max}$, a randomly chosen individual is removed from that site. The system is considered in contact with a reservoir, from where with a small (**w<<1**) probability per unit time an individual from a randomly chosen species can be assigned to a randomly chosen lattice site. This effect simulates the random fluctuations in the abundance of species. The stochastic simulation is implemented using the efficient kinetic Monte Carlo algorithm [24], and by imposing periodic boundary conditions. One starts by assigning on randomly chosen lattice sites a given number of individuals from randomly chosen species. The dynamics of the system is than straightforward. With the initially fixed probabilities we allow each individual to give birth to another individual of the same species, to die or to migrate on a nearby lattice site. We constantly verify the saturation condition for each lattice site and at each time moment we take into account the random effect of the "reservoir". After saturation on each lattice site a dynamical equilibrium sets in, and one can study the SAD. For a wide parameter range of the model, the results obtained for the SAD are quite similar to those from experimental measurements and agrees well with the results using the simulation code of McGill. The simulation code of McGill is available and described in [25].

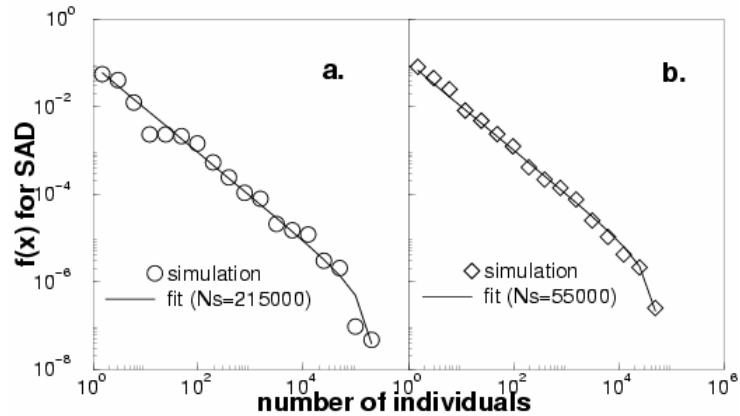

*Figure 2. Simulation results for the shape of SAD obtained by computer simulation on neutral models and fitted by equation (3). Figure 2a presents a characteristic result of our simulation code (parameters for the presented data: d/b=0.3, q/b=0.2, $N_{max}$=10000, S=400 ) and a fit with $N_s$=215000. Figure 2b presents characteristic results using the code elaborated by McGill (parameters of the presented data are: : J=$10^6$, m=0.1 and $\theta$=50 ) and a fit with $N_s$=55000.*

Fig. 2a and 2b presents simulation results for SAD on the neutral model, after the statistically stationary distribution has been reached. Figure 2a shows the SAD obtained with our simulation code when fixing the parameters as: *d/b=0.3, q/b=0.2, $N_{max}$=10000, S=400* and studying the meta-community on square of 100 lattice sites. As it is visible from the figure, equation (3) gives an excellent fit (best fit parameter: $N_s \cong 215000$). Changing the parameters of the model will not alter the qualitative shape of SAD, and equation (3) will again describe well the data. Using the simulation code of McGill and fixing his simulation parameters [25] as: $J=10^6$, $m=0.1$ and $\theta=50$, on Figure 2b we plotted the results for SAD considering the whole simulated meta-community. Choosing $N_s \cong 55000$ the shape of the obtained distribution can be well fitted. Changing the parameters of the simulation in a reasonable interval will not alter the overall shape of SAD, and equation (3) will offer again a good fit.

Accepting the form (1) for SAD, one can also derive an interesting scaling relation between the size of the most abundant species ($N_s$), the total number of individuals ($N_T$) and the number of detected species ($S_T$) in the considered habitat. From the definition of $f(x)$ it results that:

$$N_T = \int_1^{N_s} x f(x) dx = C \int_1^{N_s} \frac{N_s - x}{x} x dx$$

$$S_T = \int_1^{N_s} f(x) dx = C \int_1^{N_s} \frac{N_s - x}{x} dx \tag{4}$$

leading to the following two coupled equations:

$$N_T = C N_s (N_s - 1) - \frac{C}{2}(N_s^2 - 1)$$

$$S_T = C N_s \ln(N_s) - C(N_s - 1) \tag{5}$$

Working on relatively large habitats, one can use the $N_s >> 1$ assumption, and the coupled equation system (5) can be simplified:

$$N_T \cong \frac{C}{2} N_s^2$$

$$S_T \cong C N_s [\ln(N_s) - 1] \tag{6}$$

Eliminating from this system the normalization constant $C$, we obtain the important relation:

$$\frac{S_T N_s}{N_T [\ln(N_s) - 1]} = 2 \tag{7}$$

Computer simulation results on neutral models support the validity of the magic formula from above. On Figure 3 we show computer simulation results for different local-community sizes. Both from our simulations and from the results obtained with McGill's code it is evident that equation (7) is working, however the constant on the right side of the equation seems to be a slightly different form 2. (On the abscissa of Figure 3a, *A* represents the number of lattice sites on which SAD was constructed, and in Figure 3b *N* stands for the size of the considered local-community.) We believe that the small difference from 2 results from the crude approximation $C_1 \cong N_s$, which holds on our data only as "order of magnitude".

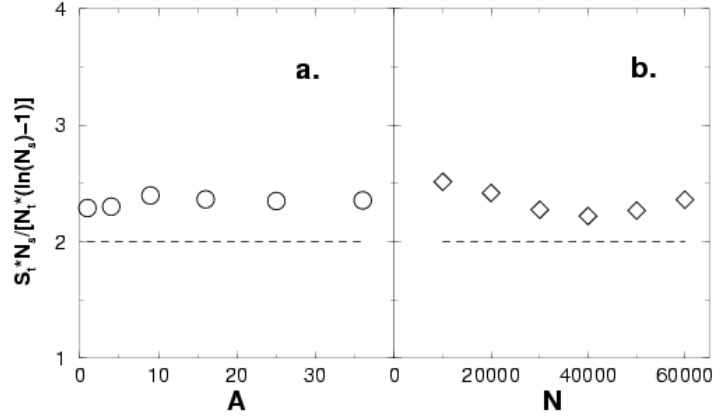

*Figure 3. Computer simulation results on the neutral model related to the validity of equation (7). Figure 3a presents results obtained using our simulation code and Figure 3b presents results obtained by using the code of McGill. The parameters of the simulations are the same as for Figure 2a and 2b, respectively.*

Increasing the size $A$ of the considered habitat, one would expect that $N_T \sim A$. It is also known that the number of species found in a habitat is scaling as a function of $A$ with a non-trivial $\alpha<1$ exponent: $S_T \sim A^\alpha$ [13]. From the above arguments one would immediately obtain, that $N_s / [\ln(N_s) - 1]$ should also follow a power-law. Using equation (7) and the $N_T \sim A$ assumption, one would also get:

$$\frac{S_T N_s}{\ln(N_s) - 1} \sim A \tag{8}$$

This scaling law can be also immediately verified. Increasing the size of the local-community, and after an ensemble average on the considered size, we can test the validity of (8).

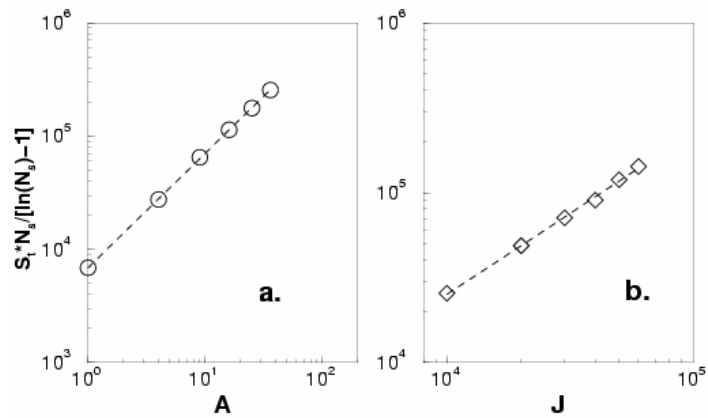

**Figure 4.** *Computer simulation results on the neutral model showing the scaling relation (8). Figure 4a presents results obtained using our simulation code and Figure 4b presents results obtained using the code of McGill. The parameters of the simulations are the same as for Figure 2a and 2b, respectively. On both figures the dashed line indicates a power-law with exponent 1.*

Both the results obtained using the simulation code of McGill and our own simulation data confirm this scaling law. The simulated data are presented on Figure 4. On this figure with a dashed line we indicated a power-law with exponent 1, and seemingly this describes well the observed scaling.

In **conclusion,** direct comparison with experimental results, and with the simulation results on neutral models confirm our conjecture (1) on the form of the SAD. The proposed distribution function for SAD leads to scaling laws confirmed by computer simulations on neutral models. We would conclude thus that the form (1) looks suitable for describing SAD in neutral meta-communities. It remains an important and interesting task to theoretically motivate the mathematical form we proposed here for fitting the SAD.

**Acknowledgements**

The present work was supported by the KPI Sapientia Foundation. We thank the many individuals who generously volunteered their time for collecting the precious data for BBS (North American Breeding Bird Survey). We gratefully acknowledge the support of the Center for Tropical Forest Science of the Smithsonian Tropical Research Institute for giving us access to the Barro Colorado Island 50 hectare Forest Dynamics Plot Data. We would like to acknowledge suggestions, criticism and helpful comments from ….